\documentclass[superscriptaddress,twocolumn,showpacs,preprintnumbers,nofootinbib,amsmath,amssymb,aps,prc,floatfix]{revtex4-1}

\usepackage{graphicx} 	
\usepackage{dcolumn}	
\usepackage{bm}		
\usepackage{amssymb}	
\usepackage{multirow}	
\usepackage{color}		
\usepackage{dcolumn}	
\usepackage{hyperref}	
\begin{document}

\title{Precision Measurement of the Neutron Beta-Decay Asymmetry}
\author{M.~P.~Mendenhall}	\affiliation{Kellogg Radiation Laboratory, California Institute of Technology, Pasadena, California 91125, USA} 
\author{R.~W.~Pattie,~Jr.}		\affiliation{Department of Physics, North Carolina State University, Raleigh, North Carolina 27695, USA} 

\author{Y.~Bagdasarova}		\affiliation{Los Alamos National Laboratory, Los Alamos, New Mexico 87545, USA} \affiliation{Department of Physics, University of Washington, Seattle, Washington 98195, USA} 
\author{D.~B.~Berguno}		\affiliation{Department of Physics, Virginia Tech, Blacksburg, Virginia 24061, USA} 
\author{L.~J.~Broussard}		\affiliation{Department of Physics, Duke University, Durham, North Carolina 27708, USA} 
\author{R.~Carr}			\affiliation{Kellogg Radiation Laboratory, California Institute of Technology, Pasadena, California 91125, USA} 
\author{S.~Currie}			\affiliation{Los Alamos National Laboratory, Los Alamos, New Mexico 87545, USA} 
\author{X.~Ding}			\affiliation{Department of Physics, Virginia Tech, Blacksburg, Virginia 24061, USA} 
\author{B.~W.~Filippone}		\affiliation{Kellogg Radiation Laboratory, California Institute of Technology, Pasadena, California 91125, USA} 
\author{A.~Garc\'ia}			\affiliation{Department of Physics, University of Washington, Seattle, Washington 98195, USA} 
\author{P.~Geltenbort}		\affiliation{Institut Laue-Langevin, 38042 Grenoble Cedex 9, France} 
\author{K.~P.~Hickerson}		\affiliation{Kellogg Radiation Laboratory, California Institute of Technology, Pasadena, California 91125, USA} \affiliation{Los Alamos National Laboratory, Los Alamos, New Mexico 87545, USA} 
\author{J.~Hoagland}		\affiliation{Department of Physics, North Carolina State University, Raleigh, North Carolina 27695, USA} 
\author{A.~T.~Holley}		\affiliation{Department of Physics, North Carolina State University, Raleigh, North Carolina 27695, USA} \affiliation{Department of Physics, Indiana University, Bloomington, Indiana 47408, USA} 
\author{R.~Hong}			\affiliation{Department of Physics, University of Washington, Seattle, Washington 98195, USA} 
\author{T.~M.~Ito}			\affiliation{Los Alamos National Laboratory, Los Alamos, New Mexico 87545, USA} 
\author{A.~Knecht}			\affiliation{Department of Physics, University of Washington, Seattle, Washington 98195, USA} 
\author{C.-Y.~Liu}			\affiliation{Department of Physics, Indiana University, Bloomington, Indiana 47408, USA} 
\author{J.~L.~Liu}			\affiliation{Kellogg Radiation Laboratory, California Institute of Technology, Pasadena, California 91125, USA}  \affiliation{Department of Physics, Shanghai Jiao Tong University, Shanghai 200240, China}
\author{M.~Makela}			\affiliation{Los Alamos National Laboratory, Los Alamos, New Mexico 87545, USA} 
\author{R.~R.~Mammei}		\affiliation{Department of Physics, Virginia Tech, Blacksburg, Virginia 24061, USA} 
\author{J.~W.~Martin}		\affiliation{Department of Physics, University of Winnipeg, Winnipeg, MB R3B 2E9, Canada} 
\author{D.~Melconian}		\affiliation{Cyclotron Institute, Texas AM University, College Station, Texas 77843, USA} 
\author{S.~D.~Moore}		\affiliation{Department of Physics, North Carolina State University, Raleigh, North Carolina 27695, USA} 
\author{C.~L.~Morris}		\affiliation{Los Alamos National Laboratory, Los Alamos, New Mexico 87545, USA} 
\author{A.~P\'erez Galv\'an}	\thanks{Currently at Physics Division, Argonne National Laboratory, Argonne, IL 60439, USA} \affiliation{Kellogg Radiation Laboratory, California Institute of Technology, Pasadena, California 91125, USA}  
\author{R.~Picker}			\affiliation{Kellogg Radiation Laboratory, California Institute of Technology, Pasadena, California 91125, USA} 
\author{M.~L.~Pitt}			\affiliation{Department of Physics, Virginia Tech, Blacksburg, Virginia 24061, USA} 
\author{B.~Plaster}			\affiliation{Department of Physics and Astronomy, University of Kentucky, Lexington, Kentucky 40506, USA} 
\author{J.~C.~Ramsey}		\affiliation{Los Alamos National Laboratory, Los Alamos, New Mexico 87545, USA} 
\author{R.~Rios}			\affiliation{Los Alamos National Laboratory, Los Alamos, New Mexico 87545, USA} \affiliation{Department of Physics, Idaho State University, Pocatello, Idaho 83209, USA} 
\author{A.~Saunders}		\affiliation{Los Alamos National Laboratory, Los Alamos, New Mexico 87545, USA} 
\author{S.~J.~Seestrom}		\affiliation{Los Alamos National Laboratory, Los Alamos, New Mexico 87545, USA}
\author{E.~I.~Sharapov}		\affiliation{Joint Institute for Nuclear Research, 141980, Dubna, Russia} 
\author{W.~E.~Sondheim}	\affiliation{Los Alamos National Laboratory, Los Alamos, New Mexico 87545, USA} 
\author{E.~Tatar}			\affiliation{Department of Physics, Idaho State University, Pocatello, Idaho 83209, USA} 
\author{R.~B.~Vogelaar}		\affiliation{Department of Physics, Virginia Tech, Blacksburg, Virginia 24061, USA} 
\author{B.~VornDick}		\affiliation{Department of Physics, North Carolina State University, Raleigh, North Carolina 27695, USA} 
\author{C.~Wrede}			\thanks{Currently at Department of Physics and Astronomy and National Superconducting Cyclotron Laboratory, Michigan State University, East Lansing, Michigan 48824, USA} \affiliation{Department of Physics, University of Washington, Seattle, Washington 98195, USA}  
\author{A.~R.~Young}		\affiliation{Department of Physics, North Carolina State University, Raleigh, North Carolina 27695, USA} 
\author{B.~A.~Zeck}			\affiliation{Department of Physics, North Carolina State University, Raleigh, North Carolina 27695, USA} \affiliation{Los Alamos National Laboratory, Los Alamos, New Mexico 87545, USA} 

\collaboration{UCNA Collaboration}
\date{\today} 

\begin{abstract}
A new measurement of the neutron $\beta$-decay asymmetry $A_0$ has been carried out by the UCNA collaboration
	using polarized ultracold neutrons (UCN) from the solid deuterium UCN source at the Los Alamos Neutron Science Center (LANSCE).
Improvements in the experiment have led to reductions in both statistical and systematic uncertainties leading to  $A_0 = -0.11954(55)_{\rm stat.}(98)_{\rm syst.}$,
	corresponding to the ratio of axial-vector to vector coupling $\lambda \equiv g_A/g_V = -1.2756(30)$.
\end{abstract}


\maketitle

Precision measurements of neutron $\beta$-decay are an essential ingredient in understanding the electro-weak interaction in the light quark sector. 
In particular the axial-vector weak coupling constant, $g_A$, 
	is an important input to understanding the spin and flavor structure of the nucleon~\cite{filippone02, bass05}
	and is being actively studied in detailed lattice QCD calculations~\cite{yamazaki08, bhattacharya12}.
It also plays an important role in a variety of astrophysical processes, including solar
	fusion cross sections important for energy and neutrino production in the sun~\cite{adelberger11}.
	
The angular distribution of emitted electrons from decays of a polarized neutron ensemble can be expressed as~\cite{jackson57}
\begin{equation}
	W(E)\propto 1 + \frac{v}{c} \langle P \rangle A(E) \cos\theta,
	\label{eq:asymmetry}
\end{equation}	
where $A(E)$ specifies the decay asymmetry for electron energy $E$,
	$v \equiv \beta c$ is the electron velocity,
	$\langle P \rangle$ is the mean neutron polarization,
	and $\theta$ is the angle between the neutron spin and the electron momentum.
The leading order value of $A(E)$, $A_0$, can be expressed as 
\begin{equation}
	A_0 = \frac{-2(\lambda ^2 - |\lambda|)}{1+3\lambda ^2},
	\label{eq:A0_lambda}
\end{equation}
where $\lambda \equiv g_A/g_V$ is the ratio of the vector to axial-vector weak coupling constants.
	Combining $g_A$ with independent measurements of the Fermi coupling constant $G_F$,
	the CKM matrix element $V_{ud}$, and the neutron lifetime $\tau_n$
	allows a precision test of the consistency of measured neutron $\beta$-decay observables~\cite{plaster12}. 

The UCNA (Ultra-Cold Neutron Asymmetry) experiment is the first experiment to use ultracold neutrons (UCN) in a precision measurement of neutron decay correlations.  
Following the publication of our earlier results (\cite{pattie09, liu10, plaster12}),
	the UCNA collaboration implemented a number of experimental improvements that led to reductions in both statistical and systematic uncertainties.
These improvements, described below, include enhanced UCN storage, improved electron energy reconstruction, and continuous monitoring of the magnetic field in the spectrometer. 
This refined treatment of the systematic corrections and uncertainties begins to address issues of consistency in the world data set for $A_0$.

The UCNA experiment ran in 2010 using the ``thin window geometry D'' as described in~\cite{liu10, plaster12},
	and collected a total of $20.6 \times 10^6$ $\beta$-decay events after all cuts are applied. 
We use the UCN source~\cite{saunders12} in Area B of the LANSCE.
UCN are polarized by a 6~T pre-polarizer magnet and a 7~T primary polarizer, coupled to an adiabatic fast passage (AFP) spin flipper to control the spin state~\cite{holley12}.
Upstream of the pre-polarizer magnet, a gate valve separates the UCN source from the experimental apparatus.

Polarized UCN enter the superconducting spectrometer (SCS)~\cite{scs_nim},
	and are confined in a 3 m long, 12.4 cm diameter diamond-like carbon (DLC) coated Cu tube (decay trap)
	with 0.7$\mu$m thick mylar endcaps.
The inside surface of each endcap is coated with 200 nm of Be to contain the neutrons.
A 0.96~T magnetic field is oriented parallel to the decay trap,
	along which decay electrons spiral toward one of two identical electron detector packages.
Between the decay trap and the detectors, the magnetic field expands out to 0.6~T,
	which reduces the electrons' transverse momenta and pitch angles, decreasing backscattering from the detectors. 
Each detector package consists of a $16\times16$ cm$^2$ low-pressure multiwire proportional chamber (MWPC)~\cite{mwpc_nim} backed by a $15$ cm diameter plastic scintillator,
	whose scintillation light is detected by four photomultiplier tubes (PMTs).
Each MWPC has 6 $\mu$m mylar windows at the front and back that separate the chamber gas (100 torr neopentane)
	from the spectrometer vacuum and PMT housing ($\lesssim 100$ torr N$_2$).
Cosmic-ray muon backgrounds are identified by a combination of plastic scintillator veto paddles
	and sealed Ar/ethane drift tube assemblies~\cite{drift_tube_nim} around the electron detectors.

A typical run unit consists of a background run (gate valve closed),
	a $\beta$-decay run (gate valve open),
	and a UCN depolarization run (see below). 
To partially cancel drifts in background and detector efficiency,
	we alternate the order of the $\beta$-decay and background runs,
	and organize the asymmetry measurements into octets with a spin flipper on ($+$), off ($-$)
	sequence of $+--+-++-$ or $-++-+--+$, chosen randomly.

Scintillator event triggers are formed by requiring at least 2 out of 4 PMT signals over threshold
	in either of the scintillator detectors.  
Due to the low mass of the MWPC, applying an analysis cut requiring coincidence between the MWPC and the scintillator
	rejects $> 99\%$ of external $\gamma$-ray background.
Energy deposition in the MWPC is calibrated against our Monte Carlo simulation to aid in classification of backscattering events.
Cosmic-ray muon backgrounds are measured and vetoed off-line
	by requiring coincidences between any of the muon detector components and the electron detectors. 

Electron positions at the MWPC are determined to $<$2~mm
	based on the distribution of charge on two perpendicular cathode grids in the MWPC~\cite{mwpc_nim}. 
A fiducial cut of $r<50$~mm (projected to the 0.96 T decay trap region)
	is placed on the trigger side to reduce background and to eliminate electrons that could strike the decay trap walls.  
	
The equilibrium UCN polarization that develops during each $\beta$-decay run
	is measured using the spin flipper to selectively unload polarized and depolarized UCN
	from the decay volume immediately following the $\beta$-decay run~\cite{plaster12}.
This is accomplished in two stages:
	first, the guide serving as input to the 7~T primary polarizing field is switched
	to guide neutrons towards a $^3$He UCN ``switcher detector''~\cite{morris09} $\sim$0.75 m below the beamline
	while the gate valve is simultaneously closed and proton pulses are discontinued.
This cleaning phase produces a signal in the UCN detector proportional to the number of correctly polarized UCN present in the experimental geometry
	at the end of the $\beta$-decay measurement interval.
The cleaning phase lasts 25 s in order to maximize depolarized UCN counting statistics
	in the subsequent measurement phase while still allowing the two time components
	of the cleaning phase spectrum to be resolved~\cite{plaster12}.
Following the cleaning phase, the state of the spin flipper is changed,
	preventing any remaining correctly polarized UCN in the decay trap from exiting the geometry
	and allowing incorrectly polarized UCN remaining downstream of the spin flipper
	to pass through the 7 T polarizing field and be counted.
Counting during this unloading phase is performed for $\sim$200 s
	in order to measure incorrectly polarized UCN as well as background.
The primary systematic uncertainty in these measurements comes from any remaining
	correctly polarized UCN upstream of the spin flipper at the moment its state is changed;
	these UCN are not prevented from reaching the UCN detector during the unloading phase
	and produce a background whose size is of the same order as the incorrectly polarized signal.
Correction for this \textit{reloaded} population is accomplished using \textit{ex situ} measurements (``reload'' measurements)
	in which the spin flipper state is toggled for 3~s during the middle of the cleaning phase
	in order to selectively enhance the signal from the reloaded population.
The measured polarization in the case of a spin-flipper-off $\beta$-decay run also requires correction
	for spin flipper inefficiency, which is determined using the difference between polarizations
	observed for spin-flipper-off and spin-flipper-on along with Monte Carlo calculated scaling factors.
Further small corrections for UCN populations detected in the switcher detector with low efficiency
	are estimated via Monte Carlo and are consistent with separate empirical studies of the system~\cite{holley12}.
These corrections include the effect of the primary polarizing magnet analyzing the unloaded UCN population with less than unit efficiency.
Based on the global agreement between Monte Carlo simulations and data, an uncertainty of 30\% is attributed to all polarization Monte Carlo calculations.
An analysis of our fitting procedure to the switcher detector signal during depolarization runs also contributes to the systematic error.
This includes sensitivity to the fitting intervals, along with the internal consistency of the extracted time constants.
\begin{ruledtabular} \begin{table}
	\caption{\label{P_table}Polarizations obtained from the two data sets:
		$2010_{I}$, which includes all 2010 depolarization and reload runs prior to the pump failure along with all 2009 reload runs,
		and $2010_{II}$, which includes all depolarization and reload runs obtained after the pump failure.}
	\begin{center} \begin{tabular}{clcc}
		& Data Set & $\langle P \rangle$ Polarization\\
		\hline
	 	& $2010_{I}$ flipper off & $1.001(2)_\mathrm{stat}(5)_\mathrm{sys}$ \\
	 	& $2010_{I}$ flipper on & $0.990(1)_\mathrm{stat}(5)_\mathrm{sys}$ \\
	 	& \\
	 	& $2010_{II}$ flipper off & $0.992(5)_\mathrm{stat}(8)_\mathrm{sys}$ \\
		& $2010_{II}$ flippper on & $0.988(4)_\mathrm{stat}(3)_\mathrm{sys}$ \\
	\end{tabular} \end{center}
\end{table} \end{ruledtabular}

Midway through the 2010 run, a vacuum pump failure unexpectedly vented the spectrometer, producing pinhole leaks in the MWPC windows.
For a brief period of operation before the windows were replaced,
	neopentane leaking from the wirechambers into the vacuum may have permanently contaminated the UCN guide surfaces,
	resulting in a change to the UCN transport characteristics of the system
	({\em e.g.} a 35\% reduction of UCN storage lifetime in the decay trap was observed after the pump failure). 
Since this incident potentially altered the equilibrium UCN polarization in the decay volume,
	separate polarization analyses for the periods before and after the pump failure were required.
In order to improve the statistics, and because there were no observable changes to the experimental geometry between the 2009 run cycle and the pump failure,
	the set of reloaded population measurements obtained in 2009 was combined with the $2010_{I}$ data acquired prior to the pump failure.
The polarizations determined from the ``before'' and ``after'' data sets are shown in Table~\ref{P_table}.

Reconstructed event energies $E_{\mathrm{recon}}$ are measured using the signals from the four PMTs
	attached by light guides to the scintillator disk in each detector.
The position dependence of light transport to each PMT is mapped out
by filling the spectrometer volume with neutron-activated Xenon. 
Natural isotopic abundance Xe gas is let into the volume normally containing the solid deuterium UCN source,
	and irradiated for a few minutes with the source flux of spallation neutrons to produce a variety of radioactive Xe isotopes by neutron capture.
After pumping the activated Xe out of the source volume, controlled amounts are introduced into the spectrometer volume.
By observing the decay spectrum features (mainly the 915~keV $\beta$-decay endpoint from $^{135}$Xe $J^\pi = \frac{3}{2}^+$)
	as a function of position using the MWPC, the position-dependent light transport of the beta scintillators is mapped out.
The increased statistics available from the Xe data compared to the previous method
	of mapping position dependence using neutron $\beta$-decay data
	allows for increased resolution and decreased statistical noise in the position-dependent response.

The energy response and linearity of each PMT is calibrated with conversion electron sources
	($^{139}$Ce, $^{113}$Sn, and $^{207}$Bi) inserted into the center of the decay trap at approximately weekly intervals~\cite{plaster12}.
The calibration source material is sealed between aluminized mylar foils.
Energy losses due to the sealing foils of each source were determined using a
	collimated $^{241}$Am alpha source and a silicon detector.
Energy losses to 5485.6~keV alpha particles passing through the mylar sealed source foils
	indicate an effective thickness of 9.5 $\mu$m
	(compared to the nominal 6 $\mu$m thickness specified by the manufacturer,
	likely due to the adhesives sealing the source package),
	uniform to $\lesssim$2\% over position on the foil.
Measured PMT response to the sources is calibrated
	using Monte Carlo simulations of scintillator energy deposition from all source decay modes,
	which include details of source encapsulation.

Since the data taken for the previous publication~\cite{liu10, plaster12},
	which required correction for nonlinearity in some of the PMTs' response due to damage from sparking in PMT bases run at sub-atmospheric pressures,
	the bases as well as the PMTs have been replaced.
The new PMTs (Hamamatsu R7725) show a linear response at the level of $< 1\%$.

The improved linearity and reduced uncertainty in position response and source foil energy losses
	allow an overall reduction in energy reconstruction systematic uncertainty to approximately half of the previous limit~\cite{liu10, plaster12}.
An energy reconstruction uncertainty of $\pm 0.31$\% on $A$ fully covers residual discrepancies between observed and Monte Carlo
	detector energy spectra for calibration sources and beta decay over the analysis energy window.

Variation in PMT and electronics gain is continuously monitored with a newly installed $^{207}$Bi ``pulser'' source,
	based on the concept of~\cite{morris76},
	consisting of a scintillator block containing a small amount of $^{207}$Bi
	mounted on the face of each PMT alongside the light guide from the main scintillator disk.
A high-threshold single-PMT trigger distinguishes pulser events from
	beta scintillator events which typically distribute light between several PMTs.
The $\sim$1MeV $^{207}$Bi conversion electron line provides a
	consistent peak for tracking gain changes, with sufficient statistics to measure each PMT's gain to $< 0.3\%$ over five minutes.
Over longer time scales, the $^{207}$Bi pulser signal peak was observed to drift on the order of 1\% per week
	relative to periodic calibrations with the conversion electron sources.
Longer term gain stabilization for $\beta$-decay data is implemented by fixing the neutron $\beta$-decay
	spectrum endpoint averaged over each octet ($\sim 8$ hours) of runs to the expected (Monte Carlo) value,
	while using the $^{207}$Bi pulsers to monitor and correct shorter time-scale drifts.

The majority of the $\beta$-decay events are single detector triggers. 
However, due to electron backscattering~\cite{schumann_2008, plaster12}, $\sim3$\% of the events trigger both scintillators,
	while $\sim2.5$\% are detected by both MWPCs but trigger only one of the scintillators.
In the first case, the initial direction of the electron can be determined by the relative timing of the triggers,
	while in the second case a cut based on the energy loss in the trigger side scintillator and MWPC
	yields an identification efficiency of $\sim$80\% based on a Monte Carlo simulation.

In our previous publications~\cite{pattie09, liu10, plaster12} the uniformity of the magnetic field in the decay region
	was checked with an NMR probe translated through the field with the central decay trap removed.
Thus the field could only be measured at the beginning and end of a long data-taking period,
	leading to additional systematic uncertainty due to possible variations between the measurements.
In the present data run the field is continuously monitored by an array of sixteen Hall effect sensors placed just outside of the decay trap.
This allows the field uniformity to be optimized each time the spectrometer magnet is ramped,
	minimizing field dips due to shim coils with damaged persistence heater switches
	that introduced magnetic field uncertainties in our previous studies~\cite{plaster12}.
Monte Carlo simulations using the observed field profile provide a correction to the measured asymmetry,
	similar to analytical estimates of magnetic mirroring for high pitch angle events.
	
In addition to the ambient backgrounds (measured with the UCN gate-valve closed) which are subtracted run-by-run,
	neutron captures in the vicinity of the detectors can create prompt $\gamma$'s and delayed $\beta$-decay electrons,
	generating an irreducible background in the experiment.
Observed events beyond the neutron $\beta$-decay endpoint after background subtraction,
	compared to a detailed Monte Carlo analysis of possible neutron capture mechanisms,
	are consistent with a particular combination of UCN capture on the aluminum surfaces of the detector and on the scintillator disk.
From this, a $\sim 0.025$~Hz neutron generated background spectrum is deduced in the energy range of the $\sim 25$~Hz $\beta$-decay signal,
	which is consistent with a small fraction of UCN escaping from small gaps in the UCN guides and decay trap,
	and within limits previously set in~\cite{plaster12}.
This excess contributes a correction and uncertainty to the measured asymmetry of $+0.01(2)$\%.

\begin{figure}
	\includegraphics[width=0.45\textwidth]{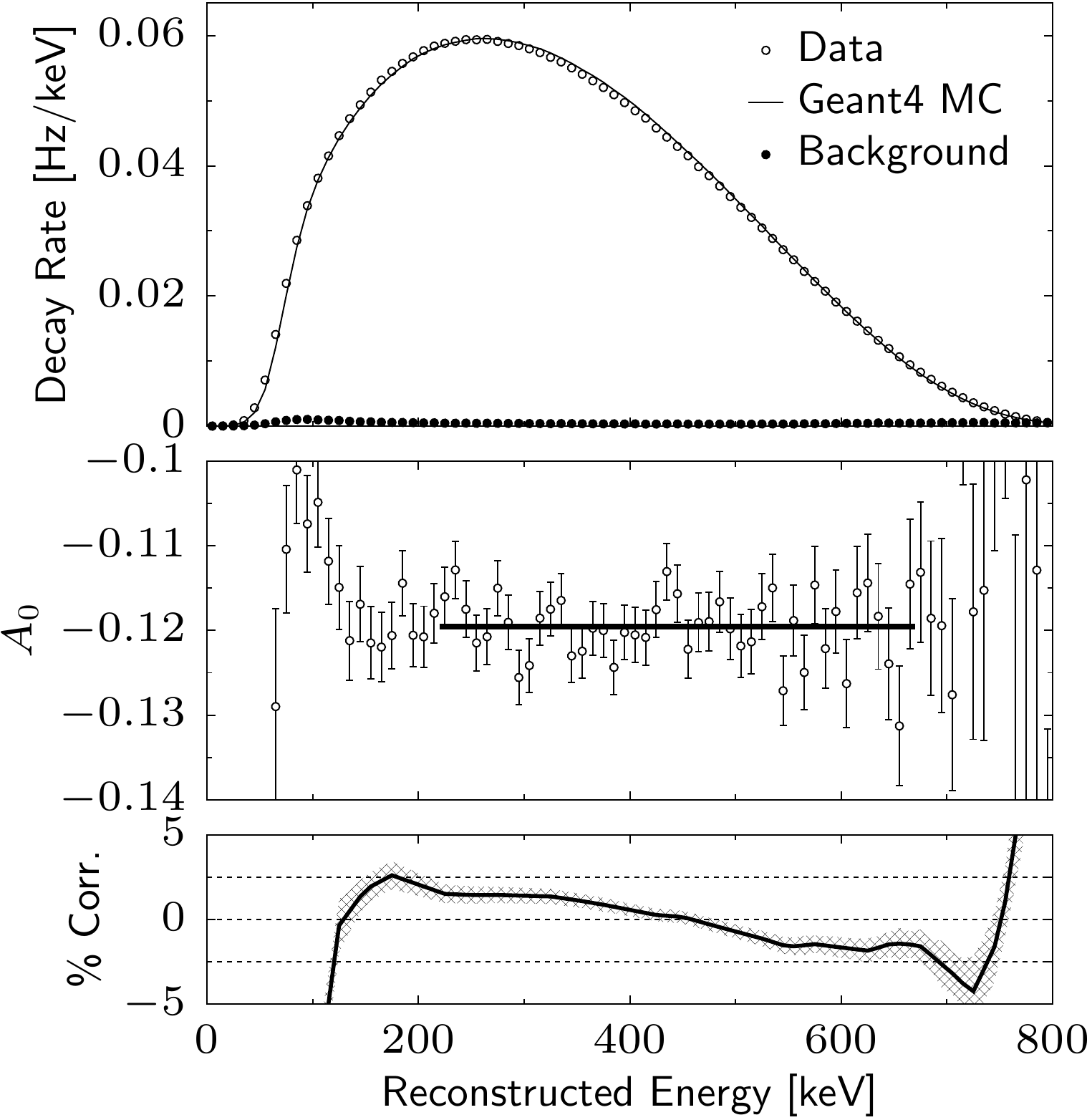}
	\caption{
  Top: background subtracted electron energy spectrum, 
  combining both detector sides and spin states, overlaid with the Monte Carlo prediction.
  The measured background spectrum is also shown.
  Middle: $A_0$ vs. $E_{\mathrm{recon}}$, shown with statistical error bars,
  and fit to a constant from 220 to 670~keV. 
  Bottom: corrections and their uncertainties (band) excluding polarization and theory contributions; positive sign indicating a larger $|A_0|$.
}
\label{fig:energy_spec}
\end{figure}

For each run, events are binned based on reconstructed energy (10~keV bins) and initial direction.
The rates in the two detectors are then computed based on the experiment live time.
We applied separate spin-dependent blinding factors to the two detector rates,
	effectively adding an unknown scaling factor to the measured asymmetry
	that was constrained to be within $1.00(5)$.
After determination of all cuts, corrections, and uncertainties, this factor was removed.
For each $\beta$-decay/background run pair,
	the background rate is subtracted from the $\beta$-decay-run rate bin by bin.
The reconstructed energy spectrum (background subtracted, averaged over the two spin states)
	is shown in Panel (a) of Fig.~\ref{fig:energy_spec}, 
	overlaid with the measured background (Signal:Background $\sim 124$ between 220 and 670~keV).
Also overlaid is the Monte Carlo predicted reconstructed energy spectrum,
	with detector response effects (energy resolution, trigger efficiency, etc.) taken into account.

In each measurement unit (octet), a ratio of count rates is constructed, leading to a ``super-ratio''
	(as defined in~\cite{pattie09}), from which the asymmetry is determined.
The final measured asymmetry is the statistical combination of all asymmetry sub-units therein. 

To extract $A_0$, we first divide the raw measured asymmetry by
	$\frac{1}{2} \beta$ in each energy bin to remove the strongest energy dependence.
As described in~\cite{pattie09, plaster12}, two scattering related effects dominate subsequent systematic corrections:
	the residual backscattering correction $\Delta_{\rm backscattering}$ and the angle effect $\Delta_{\rm angle}$.
In addition to a small correction due to incorrect identification of the
	initial electron direction for the measured electron backscatters
	(where both detectors observe the electron),
	there are corrections for backscattering from the decay trap windows
	and the front windows of the MWPC that cannot be identified experimentally.
Angle effects arise from the fact that the energy loss of an electron in the thin windows is strongly angle dependent.
Low-energy, large pitch angle electrons are more likely to fall below the scintillator threshold,
	leading to a suppression of the acceptance at large angles.
Both of these effects were evaluated with two independent Monte Carlo simulation packages:
	\textsc{Penelope}~\cite{sempau97} and \textsc{Geant4}~\cite{geant4}
	(version 4.9.5, using the ``Livermore'' low-energy EM physics model~\cite{ivanchenko11}).
The two simulations were benchmarked against the measured backscattering distributions
	for the different types of backscattering events using both neutron $\beta$-decay electrons
	and conversion-electron sources.
The resulting corrections are shown in Table~\ref{tab:sys}.
For all analysis choices (inclusion/exclusion of backscattering event types),
	the correction calculated from the two Monte Carlos agreed to within 15\%.
Based on observed differences between the simulations and the detectable backscattering data
	({\em e.g.} two scintillator triggers and two MWPC hits for single scintillator triggers),
	we assign a fractional uncertainty of 25\% to the backscattering and angle effect corrections.

Additional theoretical contributions (beyond the simple $v/c$ term) must be incorporated in order to convert
	the observable neutron beta decay asymmetry $A(E)$ to the underlying parameter $A_0$.
Recoil-order contributions to $A(E)$ were calculated within the context of the Standard Model
	according to the formalism of~\cite{bilenkii60, holstein74, wilkinson82, gardner01},
	and the radiative correction contribution was calculated according to~\cite{shann71,gluck92,gluck_pc}
	\footnote{The estimated radiative correction in~\cite{shann71}, Eq. 15, is based on an energy-independent analysis
		that integrates total counts across the whole spectrum.
		The ``Fermi function'' weighting of the spectrum towards lower energies (and lower asymmetry),
			represented by the Coulomb terms $2\pi^2 \beta^{-1}$ in \cite{shann71} Eq. 14, dominates the correction.
		For an analysis that extracts $A_0$ as a function of energy, the  bin-by-bin energy-dependent correction has the opposite sign.
		Our previous $A_0$ measurement~\cite{liu10} did not account for this.
		Updating the result with the value from table~\ref{tab:sys} modifies the result from~\cite{liu10} to $A_0 = -0.11942 \pm 0.00089^{+0.00123}_{-0.00140}$. }.
		
\newcolumntype{.}{D{.}{.}{-1}}
\begin{ruledtabular} \begin{table}
	\caption{
		\label{tab:sys}
		Summary of corrections and uncertainties as \% of $A_0$.
		``$+$'' corrections increase $|A_0|$ from the observed uncorrected value.
	}
	\addtolength{\tabcolsep}{2.5 mm}
	\begin{tabular}{l..}
		Systematic	& \multicolumn{1}{c}{corr. (\%)} & \multicolumn{1}{c}{unc. (\%)} \\
		\hline
		Polarization				& +0.67	&\pm 0.56 \\
		$\Delta_{\mathrm{backscattering}}$		& +1.36	& \pm 0.34 \\
		$\Delta_{\mathrm{angle}}$		& -1.21	& \pm 0.30 \\
		Energy reconstruction		&		& \pm0.31 \\
		Gain fluctuation			&		& \pm0.18 \\
		Field non-uniformity			& +0.06	& \pm 0.10 \\
		$\epsilon_{\mathrm{MWPC}}$	& +0.12	& \pm 0.08 \\
		Muon veto efficiency			&		& \pm0.03 \\
		UCN-induced background	& +0.01	& \pm0.02 \\
		\hline
		$\sigma_{\mathrm{statistics}}$		&		& \pm 0.46 \\
		\hline \hline
		\multicolumn{3}{c}{Theory contributions} \\
		\hline
		Recoil order~\cite{bilenkii60, holstein74, wilkinson82, gardner01}		& -1.71	& \pm 0.03 \\
		Radiative~\cite{shann71,gluck92}			& -0.10	& \pm 0.05 \\
	\end{tabular}
\end{table} \end{ruledtabular}

Applying all corrections mentioned above,
	the extracted $A_0$ is plotted against $E_{\mathrm{recon}}$
	in Panel (b) of Fig~\ref{fig:energy_spec}.
Energy-dependent corrections (backscattering and angle effects) and their uncertainty
	are indicated in the figure.
The final $A_0$ is obtained from an average over an energy range of 220 to 670~keV, 
	which was chosen, before unblinding the asymmetries,
	in order to minimize the combined statistical and systematic uncertainties.
In the 220 to 670~keV range, fitting the 10~keV binned values of $A_0$ to a constant value yields $\chi^2/{\rm ndf} = 41.7/44$ (based on statistical error bars).
The energy-averaged $A_0$ is also very stable for different energy ranges,
	remaining constant within $\pm 0.15\%$ for ranges out to 100 to 800~keV (where $\chi^2/{\rm ndf} = 68.2/69$).

\begin{figure}
	\includegraphics[width=0.47\textwidth]{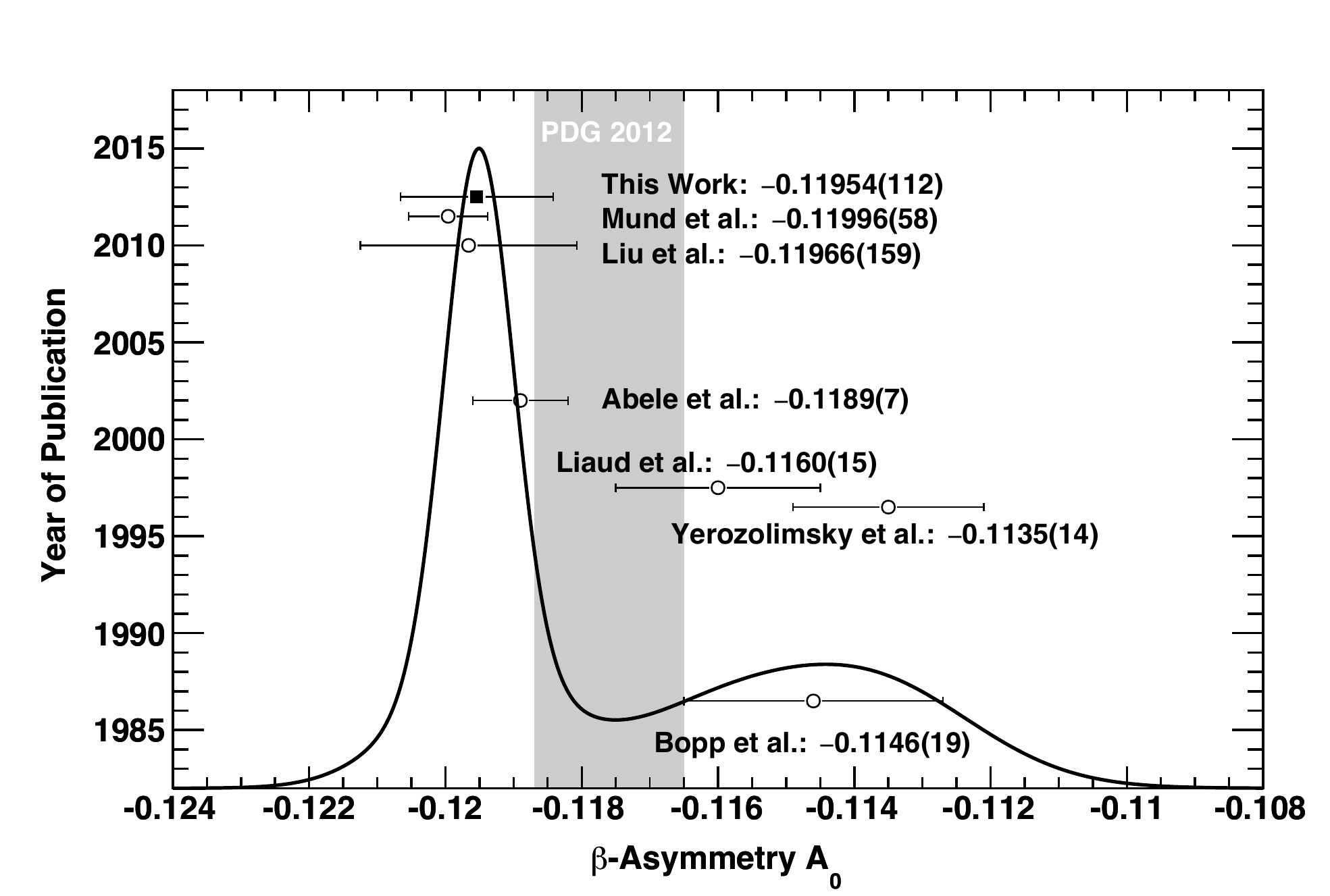}
	\caption{
	Ideogram of values for $A_0$ from this work (filled square) and recent measurements (open circles)~\cite{bopp86, yerozolimsky97, liaud97, abele02, mund12, liu10, plaster12},
		arranged by year of publication.
	To account for correlated systematic errors in sequential measurements, the ideogram (solid curve) was constructed
		using the combined result from~\cite{abele02} and~\cite{mund12} of $-0.11951(50)$ reported in~\cite{mund12},
		and the combined result of~\cite{liu10, plaster12} and this work of $-0.11956(110)$, as discussed in the text.
		The gray band indicates the PDG 2012 average value of $A_0 = -0.1176(11)$~\cite{PDG},
			which includes the results of~\cite{bopp86, yerozolimsky97, liaud97, abele02, liu10, plaster12}, but does not include~\cite{mund12} or the work reported here.
	}
\label{fig:idiogram}
\end{figure}

\begin{figure}
	\vspace{0.5cm}
	\includegraphics[width=0.40\textwidth]{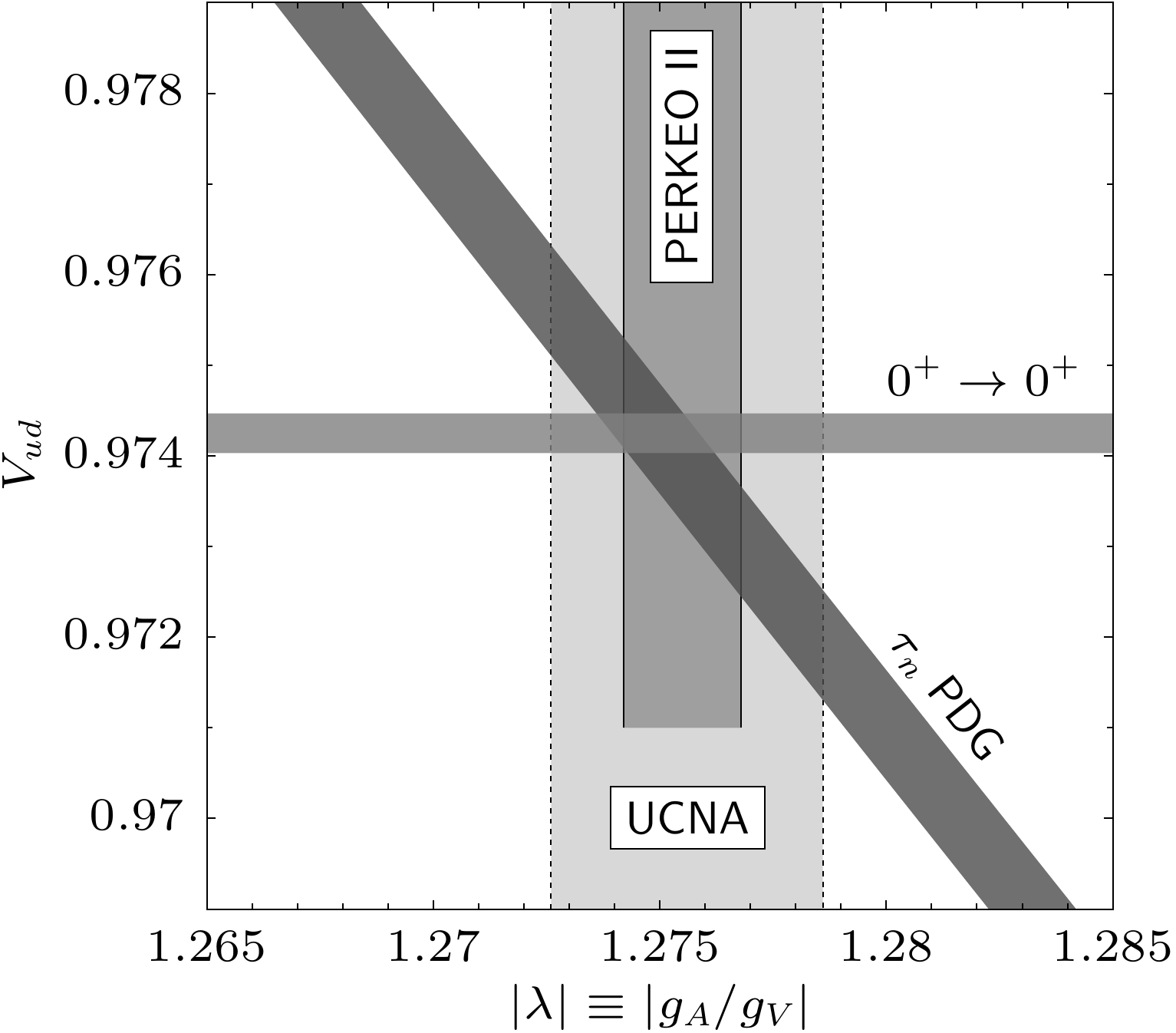}
	\caption{The light quark weak coupling $V_{\rm ud}$ vs. $\lambda$.
	$V_{\rm ud} = 0.97425(22)$ from $0^+ \rightarrow 0^+$ decays and the neutron lifetime $\tau_n = 880.1(1.1)$~s are from PDG 2012~\cite{PDG}.
	Values of $\lambda$ are the UCNA result from this paper, and the \textsc{Perkeo~II} combined result $\lambda = -1.2755(13)$ from~\cite{mund12}.}
\label{fig:phasespace}
\end{figure}

The uncertainties and systematic corrections to $A_0$ are summarized in Table~\ref{tab:sys}.
The measured result is $A_0 = -0.11954(55)_{\rm stat.}(98)_{\rm syst.}$
	where the first uncertainty is statistical and the second systematic.
Based on  Eq.\ (\ref{eq:A0_lambda}), we can also 
	determine $\lambda \equiv g_A/g_V = -1.2756(30)$.
The present result is shown in Fig.~\ref{fig:idiogram} compared with previous high precision ($\sigma_A/A < 2\%$) results. 

In summary we have measured the polarized neutron decay asymmetry with UCN resulting in a fractional precision of $< 1\%$.
When combined with our previous precision result~\cite{liu10} with the updated radiative contribution,
	 we obtain a UCNA value of $A_0 = -0.11952(110)$ and $\lambda = -1.2755(30)$.
The consistency of our results with the most recent measurements from the Perkeo collaboration~\cite{abele02, mund12},
	which have significantly smaller corrections compared to the pre-2000 results,
	may suggest that the uncertainties were under-estimated in some of these earlier experiments.
This consistency of the most recent values of $\lambda$ in the context of light quark decay parameters is shown in Fig.~\ref{fig:phasespace}.

With considerable efforts underway world-wide to improve the precision of angular
	correlations measurements sensitive to lambda using cold neutron beams~\cite{aspect_2008, perc_2008, acorn_2009, nab_2009, perkeo3_2009},
	there remains significant motivation to continue efforts to further refine corresponding measurements with UCN.

This work was supported in part by the Department of
Energy Office of Nuclear Physics (DE-FG02-08ER41557), National Science Foundation
(NSF-0855538, NSF-1205977, NSF-0653222), and
the Los Alamos National Laboratory LDRD program.
We gratefully acknowledge the support of LANSCE and
AOT divisions of Los Alamos National Lab.

\bibliography{UCNA_2010.bib}

\end{document}